\documentstyle[11pt,epsf]{article}
\begin{document}

\newcommand{\bra}{\langle}
\newcommand{\ket}{\rangle}
\def\tr{{\rm tr}\,}
\def\href#1#2{#2}

\begin{titlepage}

\begin{center}
\hfill KIAS P05014\phantom{ai\,}
\\
\hfill UOSTP 05011\phantom{a\,}
\\
\hfill hep-th/0501202
\vspace{2cm}

{\Large\bf  All Higher Genus BPS Membranes in the Plane Wave
Background}

\vspace{1.0cm}
{\large
Dongsu Bak,$\!^a$ Seok Kim$\,^b$  and Kimyeong Lee$\,^b$
}

\vspace{0.6cm}

{\it  $^a$
Physics Department, University of Seoul, Seoul 130-743, Korea
}
\vskip 0.4cm
\centerline{\it $\,^{b}$ School of Physics, Korea Institute
for Advanced Study}
\centerline{
\it Cheongriangri-Dong, Dongdaemun-Gu, Seoul 130-012, Korea}

({\tt dsbak@mach.uos.ac.kr, seok@kias.re.kr, klee@kias.re.kr})

\end{center}
\vspace{1.5cm}


We consider supermembranes in the maximally supersymmetric plane
wave geometry of  the eleven dimensions and construct complete
solutions of the continuum version of the 1/4 BPS equations. The
supermembranes may have an arbitrary number of holes and arbitrary
cross sectional shapes. In the matrix regularized version, we
solve the matrix equations for several simple cases including
fuzzy torus. In addition, we show that these solutions are
trivially generalized to 1/8 and 1/16 BPS configuations.

\vspace{3.5cm}
\begin{center}
\today
\end{center}
\end{titlepage}

\section{Introduction}

The 1/2 BPS supermembrane and their matrix regularized version in
the maximally supersymmetric plane wave geometry of the eleven
dimensions are first introduced in Ref.~\cite{Ber}. Since then the
matrix model have been studied in its various
aspects\cite{Bak}-\cite{Shin}.

In this note, we like to answer the following simple question.
In Ref.~\cite{Bak}, the 1/4 BPS equations
involving 1,2,3 directions
are found. They
are
just one mass parameter
deformation of the 1/4 BPS equations\cite{Klee,Swkim} of
supertubes\cite{Mat,Klee} in
the ordinary BFSS matrix model\cite{BFSS}.   The shape of the supertubes
involves arbitrariness; the cross sectional shape may be completely
arbitrary while preserving a quarter of  supersymmetries\cite{Kar,BO}.
Matrix solutions of
arbitrary cross sectional shapes  were constructed explicitly in
Ref.~\cite{Oht}. The mass deformation is expected  not to
ruin the arbitrariness
of supertube solutions. Thus the question is whether one can identify
the corresponding arbitrariness of  supermembranes in the
plane wave geometry.

We shall show that this is indeed possible by constructing
the most general solutions  of the
mass deformed BPS equations.
For the continuum version prior to the matrix
regularization, we are able to construct the complete solutions
of the BPS equations.
The surfaces may have an arbitrary number of holes and the cross sectional
shapes. Further, the surfaces may be either compact or noncompact.
For the matrix BPS equations,
finding general solution of the matrix
BPS equation seems very complicated. Hence we consider only the limited
cases of genus one surfaces corresponding to the fuzzy torus\cite{Mik}.

As in the case of supertubes, the general solutions may be used to identify the
degeneracy of  configurations for given charges.
One needs to fix the  energy, angular
momentum and the charges and counts the corresponding degeneracy of the
systems\cite{Mar,Hya}.
This may then be compared with the entropy of
the corresponding charged  supersymmetric black holes
as in the flat case
\cite{Dab,Sen,Rey}.
Such black hole
solution asymptotic to the plane wave geometry is not known and its
construction will be very interesting. See Ref.~\cite{Siw} for some related 
attempt to construct intersecting brane solutions asymptotic to the plane wave 
geometry.

In Section 2, we review the membrane actions and the  1/4 BPS equations.
We construct the complete solutions of the BPS equations in Section 3.
Section 4 deals with the matrix solutions. The last section comprises the
concluding remarks.

\section{Membranes in the plane wave geometry}

Our mission is to study the dynamics of  M2 branes
in the maximally supersymmetric
plane wave background,
\begin{eqnarray}
&&ds^2= -4 d x^- dx^+ -
 \left({\mu\over 6}\right)^2 \left( 4\sum^3_{i=1} x_i^2
+
\sum^9_{a=4} x_a^2
\right)
(dx^+)^2 + dx^I dx^I\nonumber\\
&& F_{+ 123}= \mu
\label{metric}
\end{eqnarray}
where the capital indices  run over $1,2, \cdots, 9$.
This geometry can be obtained from the  $AdS_7 \times S^4$
by taking the Penrose limit. The  $1,2,3$ directions are
from coordinates of $S^4$ and the remaining six
directions are related to the spatial directions of the $AdS_7$ while
one of the angular directions in $S^4$ is used
in the light cone coordinate.
The membrane action\cite{Hop,deW} for the general background is given by the
Nambu-Gotto action plus the Wess-Zumino action.
By the standard procedure\cite{Hop,deW},
this action may be
rewritten in the light cone frame as
\begin{equation}
L= L_0 + L_\mu
\end{equation}
with
\begin{eqnarray}
&&L_0={1\over 2 } \int d^2\sigma  \left( p^+ \sum_I (D_0 X_I)^2
-{1\over  p^+}
\sum_{I<J} \{X_I,X_J\}^2
+i \psi^{T} D_0 \psi + {i\over p^+} \sum_I
 \psi^{T} \gamma_I \{ X_I,\psi\} \right)\nonumber\\
&& L_\mu=
-\frac{1}{2}\int d^2\sigma \left(
\mu^2 p^+ 
\left( \frac{1}{9}\sum^3_{i=1}X_i^2 +\frac{1}{36}\sum^9_{a=4}
X_a^2\right) -{i\mu \over 3}\!\!\sum^3_{i,j,k=1}\!\!\!\! \{X_i,
X_j\} X_k \epsilon_{ijk} +{i \mu \over 4} \psi^{T} \gamma_{123}
\psi \right) \,, \label{Lag}
\end{eqnarray}
where $p^+$ is the conserved momentum related to the $x^-$ translation,
$x^+=t$ is used as the light-cone time coordinate and the Poisson bracket
is defined by
$$\{A,B\}={\partial A \over \partial \sigma_1}   {\partial B \over \partial \sigma_2}
-{\partial A \over \partial \sigma_2}   {\partial B \over \partial \sigma_1}
 \,.$$
The 16 dimensional Majonara fermions are used
for the fermionic part
and the gamma matrices are taken to be real.
One comment is that, when the spatial section of M2 brane is compact,
we call it as (an expanded) giant graviton\cite{Tou} because the quantum
numbers are matching.
The Gauss law constraint here follows from setting 
the induced world volume 
metric
\begin{equation}
g_{\,t\alpha}={\partial X^M \over \partial t\,\,} 
{\partial X^N \over \partial {\sigma^\alpha}}  
G_{MN}=0\,, 
\end{equation}
which is achieved by an appropriate worldvolume coordinate transformation. Using
the metric in (\ref{metric}), one sees that the above condition becomes
\begin{equation}
4 {\partial\over \partial {\sigma^\alpha}} X^-= \sum^9_{I=1}D_t X^I 
 {\partial\over \partial {\sigma^\alpha}}  X^I \,.
\label{gaussi}
\end{equation}
Application of $\epsilon^{\beta\alpha}{\partial/\partial {\sigma^\alpha}}$ on the both
sides leads to the Gauss law constraint. Thus the Gauss law is simply the 
integrability condition of (\ref{gaussi}). 

The matrix regularization is useful for the quantum mechanical description;
one replaces the functions on the M2 brane worldvolume 
by Hermitian $N\times N$
matrices and the Poisson brackets by commutators and so on. Namely the rules
are summarized by
\begin{eqnarray}
 \{\ \ , \  \ \} \ \  &\longleftrightarrow& \ \  -i N^2[\ \ ,\ \ ]\nonumber\\
 \int d^2\sigma \ \  &\longleftrightarrow& \ \   N \,\,\tr \nonumber\\
 X_I (\sigma) \ \  &\longleftrightarrow& \ \  {1\over N} X_I\nonumber\\
 \psi (\sigma) \ \  &\longleftrightarrow& \ \  {1\over \sqrt{N}} \psi\,.
\label{corr}
\end{eqnarray}
In addition,
we introduce a new length scale $R$ by $N/p^+$, which is identified with the
radius of $x^-$ circle.

The resulting matrix
model  Lagrangian\cite{Ber}   becomes
\begin{equation}
\tilde{L}=\tilde{L}_0 + \tilde{L}_\mu
\end{equation}
with
\begin{eqnarray}
&&2\tilde{L}_0={1\over  R}\tr \left(  \sum_I (D_0 X_I)^2
+{R^2}
\sum_{I<J} [X_I,X_J]^2
\right)  +\tr \left(i\psi^{T} D_0 \psi +  \sum_I
R \psi^{T} \gamma_I [ X_I,\psi] \right)\nonumber\\
&& 2\tilde{L}_\mu=
-\frac{\mu^2}{R}\tr \left( \frac{1}{9}\sum^3_{i=1}X_i^2
+\frac{1}{36}\sum^9_{i=4} X_i^2\right) -{2\mu i\over
3}\sum^3_{i,j,k=1}\!\!\!\!\tr X_i X_j X_k \epsilon_{ijk} -{\mu
i\over 4}\tr \psi^{T} \gamma_{123} \psi \,, \label{lag}
\end{eqnarray}
The Lagrangian $L_0$ is the same as the usual matrix model
in \cite{BFSS}. $L_\mu$ includes mass terms and the coupling
to the four form field strength background.
This matrix model has been studied in various
aspects\cite{Bak}-\cite{Shin}. Under the supersymmetry, the
fermionic field transforms as
\begin{equation} \delta \psi = \left( D_0 X_I
-\frac{i}{2}[X_I,X_J]\gamma_{IJ}+ \frac{\mu}{3}\sum_{i=1}^3
X_i\gamma_i \gamma_{123}
-\frac{\mu}{6}\sum_{i=4}^9X_i\gamma_i\gamma_{123}\right)\epsilon
\end{equation}
where $\epsilon=e^{-\frac{1}{12}\gamma_{123}t}\epsilon_0$ with a
constant spinor $\epsilon_0$.

In this note, we like to find solution of the BPS equations
constructed in Ref.\cite{Bak}. Among many others, we like to focus
first on the 1/4 BPS equations with preserved supersymmetries
$P_\mp\epsilon $ with the projection operator $P_\pm= (1\pm
\gamma_3)/2$.  The BPS equations\cite{Bak} read
\begin{eqnarray}
&& i[X_1, X_2]  + {\mu\over 3 R} X_3=0 \,, \ \
D_0 X_3 =0\nonumber\\
&& 
i[X_1, X_3]
-
{\mu\over 3R} X_2 
\pm {1\over R} D_0 X_1=0 \nonumber\\
&&  i[X_2, X_3]
+ {\mu\over 3R} X_1 
\pm {1\over R}D_0 X_2=0\,.
\label{Bpse}
\end{eqnarray}
In addition, one has to satisfy  the Gauss law
constraint,
\begin{eqnarray}
[X_1, D_0 X_1]+ [X_2, D_0 X_2]=0\,.
\end{eqnarray}
For definiteness, we shall choose the  $+$ sign projection
corresponding to the remaining
supersymmetries of
$P_-\epsilon$. The analysis below may be generalized to the other types of
BPS equations involving other directions.
Notice that these BPS equations may   be alternatively
 derived by considering
the bosonic part of the Hamiltonian. By the technique of completing
squares, one may show that the Hamiltonian is bounded by conserved
quantities,
\begin{eqnarray}
H \ \ge \  \pm \left[C_F -{\mu\over 3}J_{12} \right]\,,
\end{eqnarray}
where $C_F$ is the energy from the fundamental string tension\cite{Ban,Swkim},
\begin{eqnarray}
C_F= i \,\tr \, \left( [X_1, X_3 D_0 X_1]+ [X_2, X_3 D_0 X_2]  \right)\,,
\end{eqnarray}
and the angular momentum $J_{12}$ is defined by
\begin{eqnarray}
J_{12}={1\over R}\, \tr (X_1 D_0 X_2- X_2 D_0 X_1) \,.
\end{eqnarray}
For the compact surfaces of giant graviton, $C_F=0$  
because the representations
 are finite dimensional and the trace of commutators then vanish.
(For the continuum description, the string charge of compact surface 
is again zero because it is given by integral of total derivative.)

The above BPS equation may be simplified by taking the gauge
$A_0= R X_3$; the last two equations of (\ref{Bpse}) reduce to
\begin{eqnarray}
\dot{X}_1 +i\dot{X}_2 = -{\mu\over 3R}i (X_1+i X_2) \,.
\end{eqnarray}
Its general solution  is then given by
\begin{eqnarray}
X_1+i X_2 = e^{-{\mu\over 3R}it} (Y_1+i Y_2) \,
\label{Time}
\end{eqnarray}
where  $Y_1$ and $Y_2$ are real and time independent.  The second equation
of (\ref{Bpse}) implies that $X_3\equiv Z$ is also constant in
time. Using this solution, the full BPS equations including the Gauss law
constraint
become
\begin{eqnarray}
[Y_1, [Y_1, Z]]+ [Y_2, [Y_2, Z]] -2\left({\mu\over 3R}\right)^2
Z=0\,,\ \ \
[Y_1, Y_2]  = i{\mu\over 3R} Z \,.
\label{Bpsre}
\end{eqnarray}
These   generalize  the BPS equations associated with the matrix
description of supertubes by a mass parameter\cite{Klee,Swkim} as
noticed in Ref.~\cite{Bak}. Note that the abelian or center of
mass rotation along $X_1,X_2 $ plane decouples from the nonabelian
part without further breaking the supersymmetry.

A given 1/4 BPS configuration can be excited further by rotating
the abelian, or center of mass, part along the $X_4,X_5,...,X_9$,
while remaining BPS with less supersymmetries.   For example, we
can require that the the spinor $\epsilon_0$ satisfying the 1/4
BPS condition $\gamma_3\epsilon_0=\epsilon_0$ to satisfy an
additional compatible condition
$\gamma_{12345}\epsilon_0=\epsilon_0$, which breaks the susy
further to $1/8$ and the BPS equation along $X_4$ and $X_5$ can be
solved by $X_4+X_5 = e^{-\frac{iu}{6}t}(Y_4+iY_5)$ with constant
matrices $Y_4$ and $Y_5$   proportional to the identity matrix.
The 1/8 BPS configuration would carry nonzero angular momentum $
J_{45}$. Similarly the additional compatiable requirement
$\gamma_{12367}\epsilon_0=\epsilon_0$ breaks  susy to 1/16. The
1/16 BPS configuration satisfying $X_6+X_7 =
e^{-\frac{iu}{6}t}(Y_6+iY_7)$ where $Y_6,Y_7$ are proportional to
identity matrix would carry nonzero angular momentum $J_{67}$.
Since $\gamma_{12...9}=1$, the susy parameter $\epsilon_0$ for the
1/16 BPS also satisfies also the constraint
$\gamma_{12389}\epsilon_0 = \epsilon_0$. Thus the configuration
whose abelian part satisfying  $X_8+X_9 =
e^{-\frac{iu}{6}t}(Y_8+iY_9)$ with constant abelian $Y_8,9$ would
remain 1/16 BPS. Indeed for such a 1/16 BPS configuration, the
Hamiltonian would become
\begin{equation} H_{\frac{1}{16}} =C_F -\frac{\mu}{3}J_{12}
-\frac{\mu}{6}(J_{45}+J_{67}+J_{89}) \end{equation}
which is exactly what to expect from the BPS states from the
superalgebra.

Before closing this review, let us record the corresponding BPS
equations in the form of  the continuum description. Using the
correspondence, the BPS equations may be put into the form of
\begin{eqnarray}
&& \{X_1, X_2\}  - {\mu p^+\over 3 } X_3=0 \,, \ \
D_0 X_3 =0\nonumber\\
&&  \{X_1, X_3\}
+
{\mu p^+\over 3} X_2 
\mp  p^+ D_0 X_1=0 \nonumber\\
&&  \{X_2, X_3\}
-{\mu p^+\over 3} X_1 
\mp p^+ D_0 X_2=0\,,
\label{BpsP}
\end{eqnarray}
with the Gauss law constraint,
\begin{eqnarray}
\{X_1, D_0 X_1\}+ \{X_2, D_0 X_2\}=0\,.
\end{eqnarray}
In a similar manner, these equations may be reduced to
\begin{eqnarray}
\{ Y_1, \{Y_1, Z\}\}+ \{Y_2, \{Y_2, Z\}\} +2\left({\mu p^+\over 3}\right)^2
Z=0\,,\ \ \
\{Y_1, Y_2\}  = {\mu p^+\over 3} Z \,.
\label{BpsreP}
\end{eqnarray}

There are some known solutions of these matrix BPS equations. The
simplest ones are the 1/2 BPS vacuum solutions of zero energy and
angular momentum, which satisfies $D_0X_I=0$ and so satisfies the
SU(2) algebra and can be represented as fuzzy spheres. The 1/4 BPS
rotating ellipsoidal branes\cite{Bak} has non zero energy and
angular momentum and the corresponding configurations are
constructed  again based on the SU(2) algebra. 
The nature of the full solutions has not been   worked out.

There appeared   quite general solutions of the above continuum 
equations\cite{Mik}. (Here we shall use a completely different 
method from that of \cite{Mik}.) 
But the precise nature and implication does not seem to be understood.
As discussed
in the introduction, there are some hints from the analysis of
supertubes; there should be a huge family of solutions because the
above BPS equations generalize the BPS equation for the matrix
description of supertubes and it is now well established that
supertubes allow the arbitrary cross sectional shapes. As the BPS
configurations with less supersymmetry could build from the 1/4
BPS configurations by giving the center of mass motion, there
would be no additional degeneracy. Thus we just focus on  1/4 BPS
configurations.

Our main purpose in this note is to explore the corresponding
arbitrariness by constructing generic supermembrane solutions of
the above equations.

\section{General solutions of the continuum BPS equations}

In this section, we like to present the general solution of
the BPS equations (\ref{BpsreP})\footnote{We like to thank J.
Hoppe for pointing out a way to solve the BPS equations.}.
Let us first consider the second
BPS equation in (\ref{BpsreP}), which may be rewritten as
\begin{equation}
\sqrt{ {\rm det}\, h_{\alpha\beta}}=\left({\mu p^+\over 3}\right)\,Z
\,,
\end{equation}
where  the induced metric is defined by
$$
h_{\alpha\beta}= {\partial Y^\gamma\over \partial \sigma^\alpha}
{\partial Y^\gamma\over \partial \sigma^\beta}\,.
$$
(The Greek indices run over 1 and  2.) This can always be solved
by appropriate choice of the functions $X^\alpha(\sigma)$
for any  $Z(\sigma)$. Now noting
\begin{eqnarray}
&& \{ Y^\gamma, \{ Y^\gamma, Z\}\}=\epsilon^{\alpha\beta}
{\partial Y^\gamma\over \partial \sigma^\alpha} {\partial
Y^\delta\over \partial \sigma^\beta} {\partial\over \partial
Y^\delta}\left( \epsilon^{\lambda\kappa} {\partial Y^\gamma\over
\partial \sigma^\lambda} {\partial Y^\omega\over \partial
\sigma^\kappa} {\partial\over \partial Y^\omega} \, Z \right)
\nonumber\\
&& = \left({\mu p^+\over 3}\right)^2 Z \epsilon^{\gamma\delta}
{\partial\over \partial Y^\delta} Z \epsilon^{\gamma\omega}
{\partial\over \partial Y^\omega} Z ={1\over 2} \left({\mu
p^+\over 3}\right)^2 Z {\partial\over \partial
Y^\gamma}{\partial\over \partial Y^\gamma} Z^2
\end{eqnarray}
the first BPS equation of (\ref{BpsreP}) becomes
\begin{eqnarray}
{\partial\over \partial Y^\gamma}{\partial\over \partial Y^\gamma}
Z^2 = -4\,.
\label{lap}
\end{eqnarray}
Introducing a complex coordinate $W=Y_1+i Y_2$, the above
equation may be rewritten as
$
\partial_{\overline{W}}\partial_W Z^2 =-1\,$.
Integrating this equation once, we get
\begin{eqnarray}
\partial_W Z^2 =-\overline{W}
+ f(W)
\,,
\end{eqnarray}
where $f(W)$ is an arbitrary  function of only $W$ at this
stage.
One may take $f(W)$ as the following form,
\begin{eqnarray}
f(W)= \sum_k {a_k\over W-b_k} + g(W)
\,,
\end{eqnarray}
where $g(W)$ is analytic on the worldvolume of supermembrane.
The precise restriction of $G(W)$ will be specified further below.
One more integration leads
to the solution
\begin{eqnarray}
Z^2 + |W|^2= A + G(W) + \overline{G(W)}
+ \sum_k a_k\ln(W-b_k) + \sum_k \bar{a}_k\ln(\overline{W}-\bar{b}_k)
\,,
\end{eqnarray}
where $G(W)$ is the integral of $g(W)$ and $A$ is a real
constant. To avoid the possibility of singular surfaces,
$a_k$ should be taken to be real. Then the final form of the
solution reads\footnote{Strictly speaking, (\ref{Solution})
is the solution of (\ref{lap}) with delta-function sources at $W=b_k$. However,
the sources are not included into the worldvolume of the membrane, which defines 
the domains of Eq.~(\ref{lap}).}
\begin{eqnarray}
Z^2 + |W|^2= A + G(W) + \overline{G(W)}
+ \sum^g_{k=1} a_k\ln|W-b_k|^2
\,.
\label{Solution}
\end{eqnarray}
As mentioned before, the Gauss law constraint is the integrability condition 
for the $X^-$ equation. 
Since we solved the Gauss law constraint,  $X^-$ equation 
in (\ref{gaussi}) may be integrated.
This is ture locally  but there may remain  a global 
issue\footnote{We like to thank Andrei Mikhailov for bringing 
this issue to us.}\cite{Mik}. To see this, first note 
that Eq.~(\ref{gaussi}) may be rearranged to the form
\begin{eqnarray}
4 {\partial\over \partial Y^\alpha} X^- =
{\mu i\over 6}{\partial\over \partial Y^\alpha}
\left( G(W)-\overline{G(W)}+
\sum^g_{k=1} a_k\left[\ln(W-b_k) - \ln(\overline{W}-\bar{b}_k)\right]
\right)
\,,
\label{gaussb}
\end{eqnarray}
where we used the BPS equations, the solution (\ref{Solution}) 
and the fact
$\partial G/\partial X_1 = -i \partial G/\partial X_2 $. 
The solution becomes then\cite{Mik}
\begin{eqnarray}
{X^-\over R\,\,\,}= -{\mu\over 12 R}\left ( C(t)+ {\rm Im} (G) +
\sum^g_{k=1} a_k\theta_k
\right)
\,,
\label{gaussc}
\end{eqnarray}
where the angle variable  $\theta_k$ is defined by 
$
[\ln(W-b_k)-\ln( \overline{W}-\bar{b}_k)]/(2i)$ and $C(t)$ a
function of time only. For the global 
issue, only the logarithmic parts are relevant. 
In order to have a surface that is closed 
in the space including the $x^-$ 
direction, the coefficient   
$\alpha_k ={\mu\over 12 R}a_k$ has to be a rational
number. For $g=1$ case, $\theta_1$ and $x^-$ space has a topology of torus.
The above solution represents a curve in this space, which is only closed when
$\alpha_1$ is fractional. This argument may be generalized to the higher genus 
cases too. 

One may wonder what is happening after the matrix regularization.
For the finite dimensional matrix configurations, such a topological 
information will be lost completely and one finds no restriction for the 
coefficient $a_k$. For the noncompact infinite dimensional configurations,
one may consider the case of supertubes corresponding to $\mu=0$ limit. 
The logarithmic behaviors in the above is closely related to membrane 
configuration pulled out by the fundamental strings\cite{Oht,HHy,Sad}.  
The winding corresponding to the fundamental string charges was found to be
fractional  after taking care of the quantization of the dynamical 
modes of supertubes\cite{Hya}. Further investigation of this issue for the case 
of the matrix model will be of interest.

Let us now discuss about details of solutions.
The choice of  $G(W)= a_k=0$ corresponds to the spherical membrane
$Z^2+ |W|^2= l^2$. The arbitrary deformations of the sphere by
introducing $G(W)$ will be also solutions. 
For instance, the ellipsoid in Ref.~\cite{Bak} is given with the choice
of $G(W)=c\, W^2$ with $|c| < 1$. For small deformation the sphere topology remains.
For large deformation, the supermembrane may even become noncompact. This happens
for instance choosing $|c| > 1$.

\begin{figure}[htb] \epsfxsize=3.2in
\hspace{0.8in}\epsffile{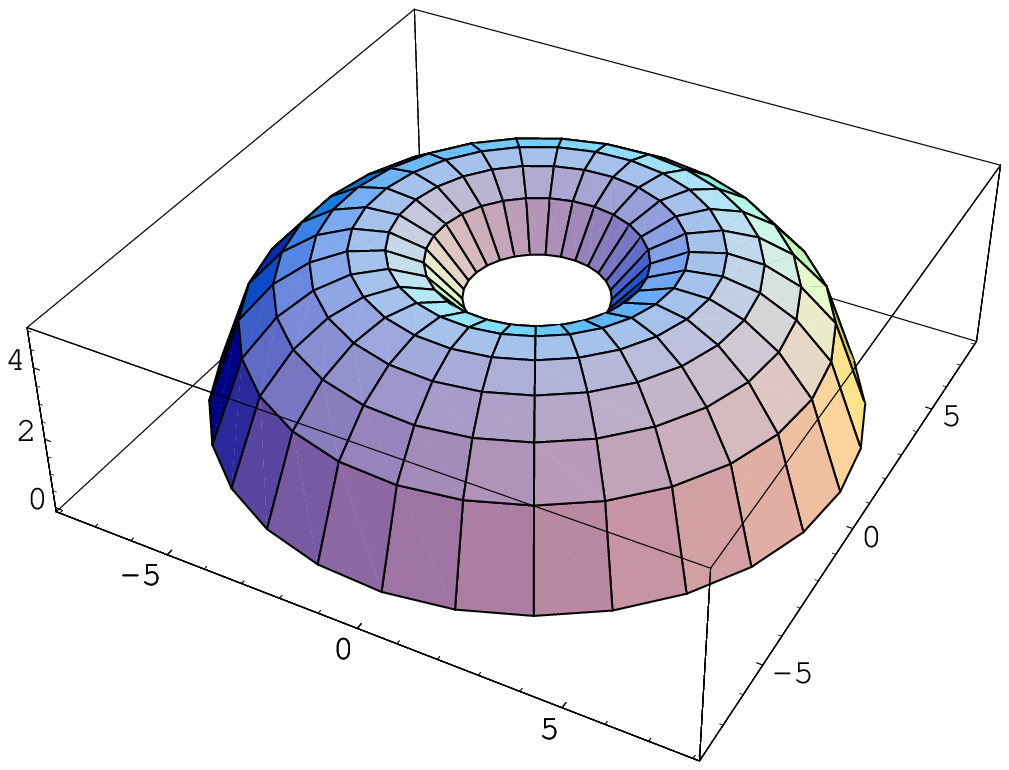}
\vspace{.2in}
\\
{\small Figure~1:~We illustrate the genus one supermembrane
 drawn for $z \ge 0$.
}
\end{figure}

The Hamiltonian and the angular momentum defined by
are conserved and, using the BPS equations, one may show
that
\begin{eqnarray}
H=-{\mu\over 3}J={\mu^2 p^+\over 9}\int d^2 \sigma (Y_1^2+Y_2^2-2 Z^2)
\,,
\end{eqnarray}
for the compact surfaces. Only noncompact surfaces may carry  nonvanishing
fundamental string charges.
For the sphere, one may easily see that
the energy as well as the angular momentum vanish. The supermembrane of
ellipsoidal shape has nonvanishing energy and angular momentum.

Let us turn to the case of $g=1$. Taking $G(W)=0$, one has
\begin{eqnarray}
Z^2 + |W|^2 - a \ln (|W|^2/a)=l^2 + a
\,,
\label{one}
\end{eqnarray}
with $a > 0$.
In Figure~1, we depict the shape of the  genus one
surface. 
When $l=0$, the surface becomes a
circle of radius $\sqrt{a}$ in the $Z=0$ plane. For finite $l$,
the surface has the shape of  doughnut, which is genus one.

All the higher genus solution may be constructed in a similar
manner. The positive integer
$g$ in (\ref{Solution})
counts the  genus number, i.e. the number of holes.
Figure~2  illustrates a genus two surface
described by
\begin{eqnarray}
Z^2 + |W|^2 - a \ln |W-b|^2 - a \ln |W+b|^2 =A
\,,
\end{eqnarray}
where again  $a > 0$ and we take $b$ to be real for simplicity.

Noncompact surfaces are also possible. Whenever $a_k$ is negative,
the resulting surface is infinitely extended in the $\pm \, z$ directions
at $W= b_k$. Figure~3 shows a simple such case with the rotational
symmetry around z-axis; it is described by the equation,
\begin{eqnarray}
Z^2 + |W|^2 + \tilde{a} \ln (|W|^2/\tilde{a})=l^2 + \tilde{a}
\,,
\end{eqnarray}
with $\tilde{a} > 0$. If one introduces many logarithms with
negative $a_k$, the supermembranes has correspondingly many
spikes in the $\pm \, z$ directions.

\begin{figure}[bht] \epsfxsize=3.2in
\hspace{0.8in}\epsffile{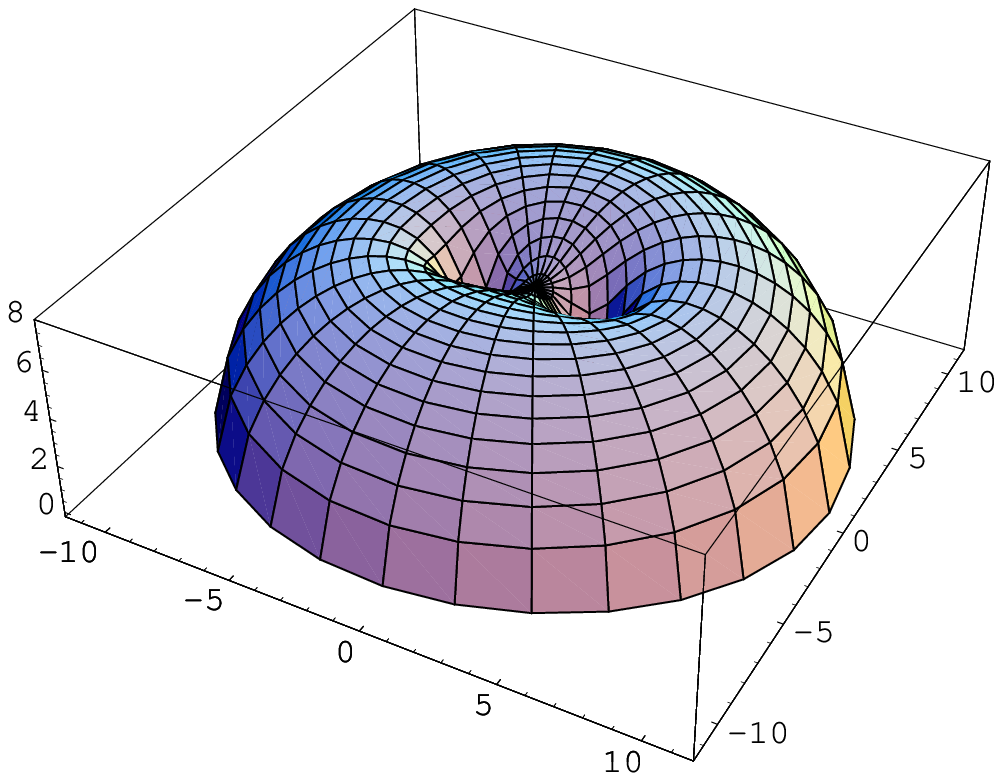}
\vspace{.2in}
\\
{\small Figure~2:~The shape of  genus two supermembrane drawn for $z\ge 0$.
}
\end{figure}

There is again arbitrariness of the shape by changing the analytic
function $G(W)$.  By an appropriate scaling of the variables in the
limit $\mu \rightarrow 0$, one may recover the supertube
solutions\cite{Bak,Swkim,Oht} from the one in Figure 3 and its
deformations.
One example of such scaling limit is to take 
$X_1\rightarrow X_1$, $X_2\rightarrow X_2$, 
$Z\rightarrow Z/\sqrt{\mu}$ as
$\mu\rightarrow 0$. One can see that the BPS equations for
supertubes are recovered in this scaling limit.

\begin{figure}[htb] \epsfxsize=1.8in
\hspace{1.3in}\epsffile{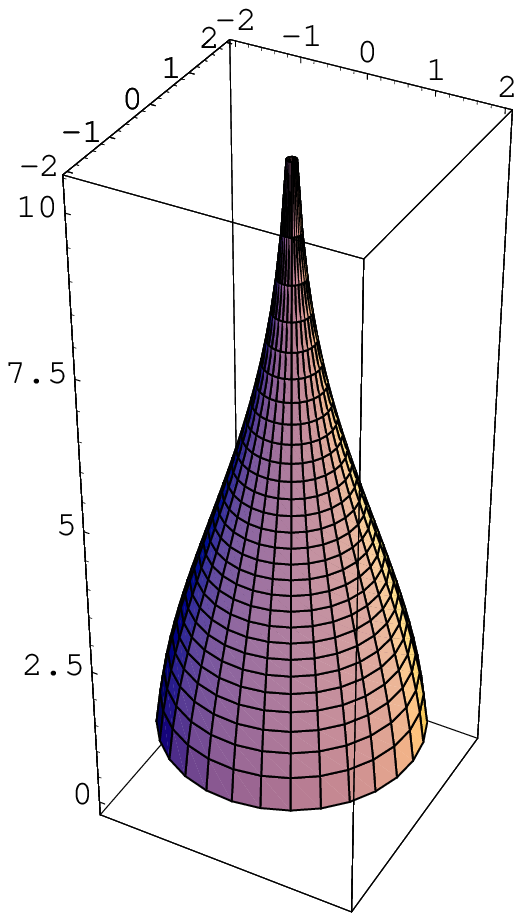}
\vspace{.2in}
\\
{\small Figure~3:~We illustrate a supermembrane of tubular shape
for $z \ge 0$.}
\end{figure}

Finally, let us discuss the precise restriction on the possible choice of
$G(W)$. Let the domain of supermembrane be the projection of
the worldvolume of supermembrane to the $W$ plane. Then $G(W)$ should be analytic
in the closure of the domain of supermembrane. Singularities of the type
$(W-b)^{-n}$ with $n>0$ are not allowed within the closure of the domain
in order to avoid singular surfaces.

\section{Solutions of matrix BPS equations}

In the previous section, we presented the general solution of our
continuum BPS equations. We now like to analyze the matrix BPS equations
in (\ref{Bpsre}). As discussed in Ref.~\cite{Mik}, the matrix BPS equations are
solved by the following algebra,
\begin{eqnarray}
&& [W_+, W_-]=2\left({\mu\over 3R}\right) Z \nonumber\\
&&  [Z, W_-]= - \left({\mu\over 3R}\right) W_- + f(W_+)\,,
\end{eqnarray}
where $W_+= Y_1+i Y_2$ is now a matrix and $f(W_+)$ is a 
function of only $W_+$. We do not know how to solve this algebra
generically. Instead, we like to solve for the case of fuzzy
torus, which is described by the above algebra with the choice of
$$f(Y_+)= \left({\mu\over 3R}\right)^3 {a\over  Y_+}\,.$$
For the simplicity of the expression, let us introduce
the following scaled variables,
\begin{eqnarray}
L_1= \left({3R\over \mu}\right) \, Y_1\,,\ \
L_2= \left({3R\over \mu}\right)\, Y_2 \,,\ \
L_3 = \left({3R\over \mu}\right) \, Z \,,
\end{eqnarray}
and $L_+= L_1+i L_2$.

For the $n$ dimensional representation with $L_3$ diagonalized,
the algebra turns into a set of simple
algebraic equations  if one takes an
ansatz,
\begin{eqnarray}
L_+=\sum^n_{i=1} b_i\,|i\rangle \langle i+1|\,,
\label{ansat}
\end{eqnarray}
where we define $|n+1\rangle=|1\rangle$. The ansatz here is partly motivated
by the consideration of the rotational symmetry.

\begin{figure}[htb] \epsfxsize=4.5in
\hspace{0.8in}\epsffile{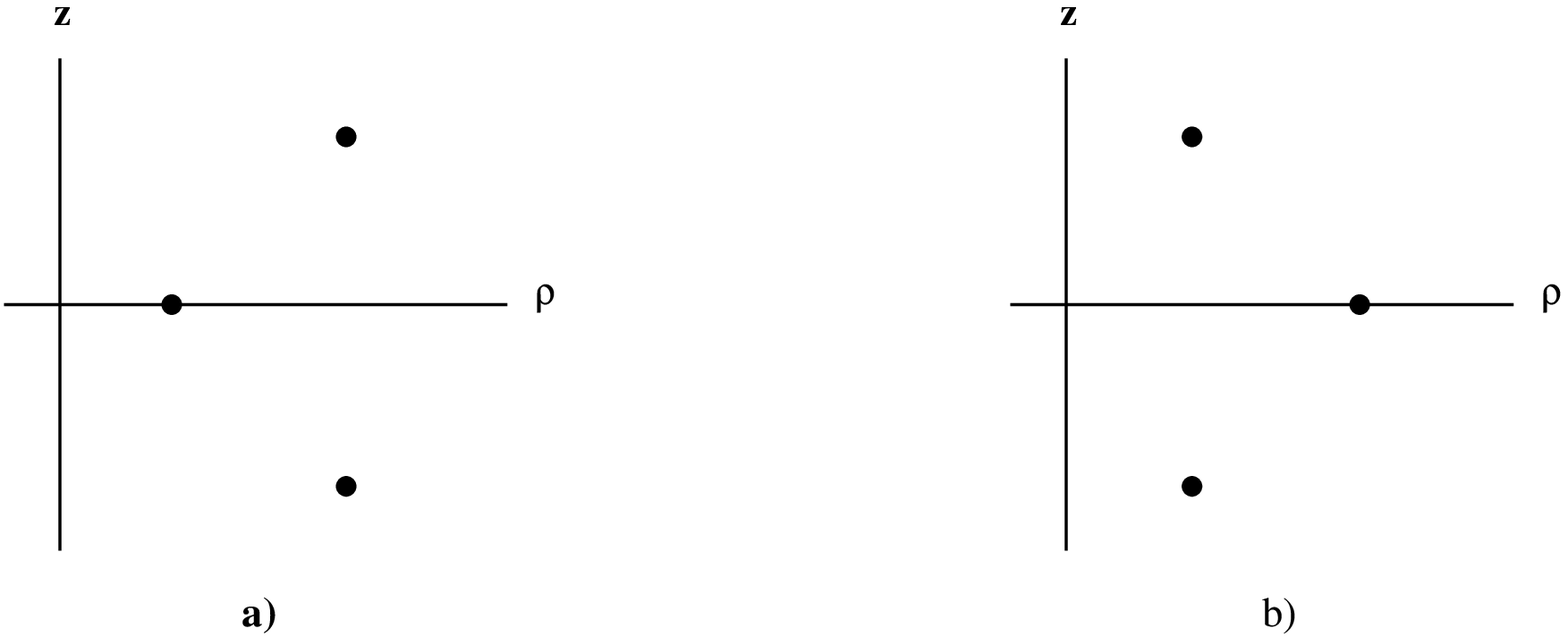}
\vspace{.2in}
\\
{\small Figure~4:~The locations of the eigenvalues in $z-\rho$ plane;
a) is for the upper sign or for the lower sign with $ 0 \le a \le 2/3$
and b) for the lower sign with $  2/3 \le a \le 3/4$.
}
\end{figure}

Using $(L_+)^{-1}=
\sum^n_{i=1} b_i^{-1}\,|i+1\rangle \langle i|$, the BPS
equations become
\begin{eqnarray}
&& 2 z_i= |b_{i-1}|^2 -|b_i|^2\,, \nonumber\\
&&  z_{i+1}- z_i= 1- a/|b_i|^2\,,
\end{eqnarray}
where $z_i$ is the diagonal entry of $L_3$. Here we use the convention
$z_{n+1}=z_1$,  $z_0=z_n$ and $b_0=b_n$.
For $n=2$, no nontrivial representation with $L_3\neq 0$.
For $L_3=0$, one may solve the problem for general $n$. Simply one has
 $|b_k|= \sqrt{a}$ for all $k$. $L_+$ can be diagonalized with eigenvalues
$\sqrt{a}e^{{2\pi k\over n}i}$ with $k=1,2,\cdots, n$.
This is not the most general solution with $Z=0$;
it may be found by discarding the ansatz (\ref{ansat}) and directly solving
the algebra with
$Z=0$.
 In the diagonalized basis
of $L_+$, the only condition is that the absolute of the eigenvalues should
be $\sqrt{a}$, which agrees with the continuum case. For the solutions,
$H=-\mu J /3= {n a\mu^2/(9 R)}$, so they carry a nonvanishing angular
momentum.

For $n=3$, we have just considered the case where all $|b_k|^2$'s are
the same. In the remaining cases, at least two of $|b_k|^2$ should
be the same. Without loss of generality, we let $|b_1|^2=|b_2^2|$ using
the cyclic property of our ansatz. The solutions are given by
\begin{eqnarray}
&& |b_1|^2=|b_2|^2= {1\pm \sqrt{1-4a/3}}\,, \nonumber\\
&&  |b_3|^2={1\mp \sqrt{1-4a/3}\over 2 }\,,
\end{eqnarray}
where $a$ is constrained by $ 0
\le a \le {3\over 4}$.
The corresponding $L_3$ is given by
\begin{eqnarray}
L_3=\left[
\begin{array}{ccc}
 -{1\pm 3\sqrt{1-4a/3}\over 4} & 0 & 0 \\
 0 & 0 &  0 \\
  0 & 0 & {1\pm 3\sqrt{1-4a/3}\over 4}
\end{array}\right]
\end{eqnarray}
In this solution, $\rho^2$ defined by $L_1^2+L_2^2$ is also diagonal, which
reflects the rotational symmetry around $z$ axis. Its expression
 is given by
\begin{eqnarray}
\rho^2=\left[
\begin{array}{ccc}
 {3\pm \sqrt{1-4a/3}\over 4} & 0 & 0 \\
 0 & 1\pm \sqrt{1-4a/3} &  0 \\
  0 & 0 & {3\pm \sqrt{1-4a/3}\over 4}
\end{array}\right]
\end{eqnarray}
The locations of the eigenvalues in $z-\rho$ plane are depicted in Figure~3.
{}From the Figure~4, one can see that
the crude feature of genus one surface emerges
though the topology here is not well defined due to  fuzziness.
The Hamiltonian for this solution is evaluated as
\begin{eqnarray}
H=- \mu J/3={\mu^2\over 36R } \left({5+6a\pm 3 \sqrt{1-4a/3}}\right)\,.
\end{eqnarray}

\begin{figure}[hbt] \epsfxsize=4.5in
\hspace{0.8in}\epsffile{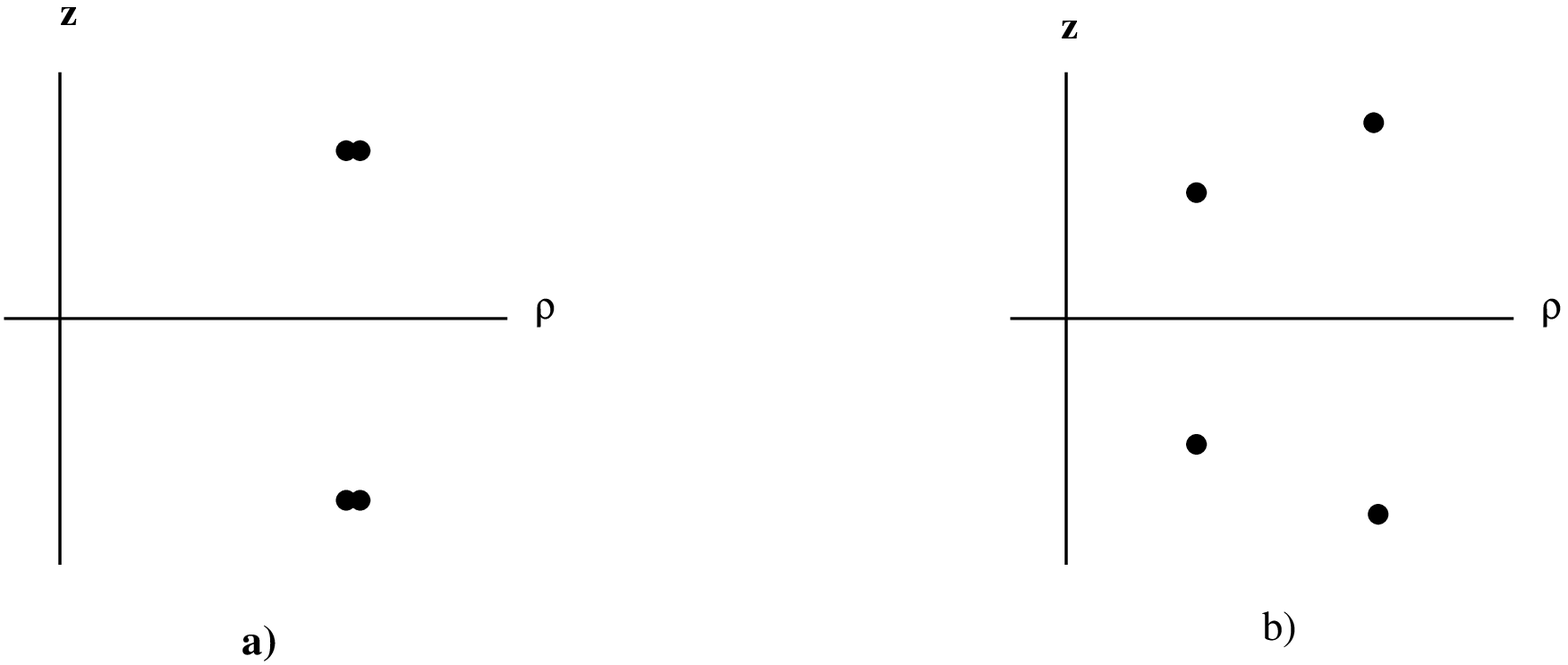}
\vspace{.2in}
\\
{\small Figure~5:~The locations of the eigenvalues in $z-\rho$ plane with
$4\times 4$ matrices;
a) is for the case where $|b_1|^2=|b_2|^2$ and
$|b_3|^2=|b_4|^2$
and b) for the case of $|b_1|^2=|b_3|^2$.
}
\end{figure}

Let us now move to the cases of $4\times 4 $ matrices. There are three
classes of solutions. The first is the
one with the same  $|b_k|^2$'s for all $k$ and $L_3=0$,
which is already described
in the above. The second is the case where $|b_1|^2=|b_2|^2$ and
$|b_3|^2=|b_4|^2$ up to any permutation
of the entries. The  solution of the
algebraic equations is found as
$|b_1|^2=|b_2|^2={1\pm \sqrt{1-a}}$ and
$|b_3|^2=|b_4|^2={1\mp \sqrt{1-a}}$ where
$0\le  a\le 1$.
The eigenvalue distributions of $L_3$ and $\rho^2$ are drawn
in Figure~5a.
The corresponding $L_3$ is given by
\begin{eqnarray}
L_3=\left[
\begin{array}{cccc}
 \mp\sqrt{1-a} & 0 & 0 & 0 \\
 0 & 0 &  0  & 0 \\
  0 & 0 & \pm\sqrt{1-a} & 0\\
  0 & 0 &  0  & 0
\end{array}\right]
\end{eqnarray}
and
\begin{eqnarray}
\rho^2=\left[
\begin{array}{cccc}
 1  & 0 & 0 & 0 \\
 0 &  1\mp\sqrt{1-a} &  0  & 0 \\
  0 & 0 & 1& 0\\
  0 & 0 &  0  & 1\mp\sqrt{1-a}
\end{array}\right]
\end{eqnarray}

The last one is the case where only one pair of entries
 has the same values, e.g.
 $|b_1|^2=|b_3|^2$. In this case,
$|b_1|^2=|b_3|^2= (3\pm \sqrt{9-8a})/2$,  
$|b_2|^2|b_4|^2= a$ and $ |b_2|^2+|b_4|^2=
(5\pm \sqrt{9-8a})/2$. One finds then
$z_1=-z_4$ and $z_2=-z_3$. The eigenvalue distribution is depicted
in Figure~5b.

In these solutions the parameter $a$ is bounded from above and below. Let us note 
that, for the case of fuzzy sphere, the size of the sphere is determined once the 
dimension of representations  is given. The above restriction on the parameter
$a$ for a given dimension of matrices may be understood in a similar context. 
The parameter $a$ here is related to the size of the hole of the genus one 
surface.  

One can consider the infinite case $N$ but with discrete matrix
case. The 1/4 BPS solution represented by this case cannot be
confined in a finite space.  The solution becomes simplified when
there is a rotational symmetry, and the generalization of the
above ansatz leads to a recursion relation between coefficient.
This would be the matrix generalization of the supertube in the
flat space~\cite{Klee}.  As the general characteristics of these
solutions have similar to those of  the generalized supertube
found in the supermembrane solution discussed in the previous
section, we will not present the detail of the solution. Finally we like to mention 
again that the cotinuum limit of the previous section may be recovered by 
taking the large $N$ limit while keeping $X_I/N$ finite.

\section{Conclusion}

In this note, we consider the  BPS supermembrane configurations
 in the maximally supersymmetric plane wave background in
 eleven dimensions. The 1/2 BPS configurations are well known
 spherical giant gravitons only with $p^+$ and carries zero
 angular momentum. For the 1/4 BPS equations, the complete solutions
 of supermembranes are constructed in full generality. They are characterized by
arbitrary number of holes as well as the arbitrary cross sectional
shapes. They can be either compact or  noncompact. For the matrix
regularized version, we constructed several solutions focusing on
the case of the fuzzy torus. The 1/4 BPS supermembranes are in
three space defined by   the nonvanishing four form tensor field
strength. Additional angular momentum along other directions by
rotating the center of mass reduces the invariant supersymmetry to
1/8 and 1/16.

The construction may used to the counting of the degeneracy of the
fixed energy and angular momentum  sectors for a given charges as
done in Ref.~\cite{Hya}. The construction of the corresponding
black hole solutions will be very interesting. Also the
characteristics of the less supersymmetric supermembrane suggests
the  direction to pursue for similar BPS configurations in the
$AdS^7\times S^4$ or $AdS^4\times S^7$.

\noindent{\large\bf Acknowledgment}
We would like to thank J. Hoppe, Andrei Mikhailov,  Piljin Yi
and Ho-Ung Yee for 
the helpful discussions. D.B. is supported in part by KOSEF ABRL
R14-2003-012-01002-0 and D.B and K.L. are supported in part by
KOSEF R01-2003-000-10319-0. D.B. also likes to thank the warm
hospitality of KIAS where part of this work was done.



\begin{thebibliography}{99}


\bibitem{Ber}
D.~Berenstein, J.~M.~Maldacena and H.~Nastase,
JHEP {\bf 0204}, 013 (2002)
[arXiv:hep-th/0202021].



\bibitem{Bak}
D.~Bak,
Phys.\ Rev.\ D {\bf 67}, 045017 (2003)
[arXiv:hep-th/0204033].







\bibitem{Das}
K.~Dasgupta, M.~M.~Sheikh-Jabbari and M.~Van Raamsdonk,
JHEP {\bf 0205}, 056 (2002)
[arXiv:hep-th/0205185].





\bibitem{Mik}
A.~Mikhailov,
arXiv:hep-th/0208077.



\bibitem{Park}
J.~H.~Park,
JHEP {\bf 0210}, 032 (2002)
[arXiv:hep-th/0208161].

\bibitem{Lee}
K.~M.~Lee,
Phys.\ Lett.\ B {\bf 549}, 213 (2002)
[arXiv:hep-th/0209009].



\bibitem{Hyun}
S.~j.~Hyun and H.~j.~Shin,
Phys.\ Lett.\ B {\bf 543}, 115 (2002)
[arXiv:hep-th/0206090].


\bibitem{Kim}
N.~w.~Kim and J.~Plefka,
Nucl.\ Phys.\ B {\bf 643}, 31 (2002)
[arXiv:hep-th/0207034];
K.~Dasgupta, M.~M.~Sheikh-Jabbari and M.~Van Raamsdonk,
JHEP {\bf 0209}, 021 (2002)
[arXiv:hep-th/0207050].
N.~Kim and J.~H.~Park,
Phys.\ Rev.\ D {\bf 66}, 106007 (2002)
[arXiv:hep-th/0207061].

\bibitem{Sug}
K.~Sugiyama and K.~Yoshida,
Phys.\ Lett.\ B {\bf 546}, 143 (2002)
[arXiv:hep-th/0206132];
K.~Sugiyama and K.~Yoshida,
Nucl.\ Phys.\ B {\bf 644}, 113 (2002)
[arXiv:hep-th/0206070];
N.~Nakayama, K.~Sugiyama and K.~Yoshida,
Phys.\ Rev.\ D {\bf 68}, 026001 (2003)
[arXiv:hep-th/0209081];
K.~Sugiyama and K.~Yoshida,
Phys.\ Rev.\ D {\bf 66}, 085022 (2002)
[arXiv:hep-th/0207190];

\bibitem{Shin}
H.~Shin and K.~Yoshida,
Nucl.\ Phys.\ B {\bf 679}, 99 (2004)
[arXiv:hep-th/0309258];
H.~Shin and K.~Yoshida,
arXiv:hep-th/0409045.

\bibitem{Klee}
D.~Bak and K.~Lee,
Phys.\ Lett.\ B {\bf 509}, 168 (2001)
[hep-th/0103148].


\bibitem{Swkim}
D.~Bak and S.~W.~Kim,
Nucl.\ Phys.\ B {\bf 622}, 95 (2002)
[arXiv:hep-th/0108207];


\bibitem{Mat}
D.~Mateos and P.~K.~Townsend,
``Supertubes,''
Phys.\ Rev.\ Lett.\  {\bf 87}, 011602 (2001)
[hep-th/0103030].



\bibitem{BFSS}
T.~Banks, W.~Fischler, S.~H.~Shenker and L.~Susskind,
Phys.\ Rev.\ D {\bf 55}, 5112 (1997)
[hep-th/9610043].






\bibitem{Kar}
D.~Bak and A.~Karch,
Nucl.\ Phys.\ B {\bf 626}, 165 (2002)
[arXiv:hep-th/0110039];
D.~Mateos, S.~Ng and P.~K.~Townsend,
JHEP {\bf 0203}, 016 (2002)
[arXiv:hep-th/0112054].

\bibitem{BO}
D.~Bak and N.~Ohta,
Phys.\ Lett.\ B {\bf 527}, 131 (2002)
[arXiv:hep-th/0112034].



\bibitem{Oht}
D.~Bak, N.~Ohta and M.~M.~Sheikh-Jabbari,
JHEP {\bf 0209}, 048 (2002)
[arXiv:hep-th/0205265].







\bibitem{Mar}
B.~C.~Palmer and D.~Marolf,
JHEP {\bf 0406}, 028 (2004)
[arXiv:hep-th/0403025].

\bibitem{Hya}
D.~Bak, Y.~Hyakutake and N.~Ohta,
Nucl.\ Phys.\ B {\bf 696}, 251 (2004)
[arXiv:hep-th/0404104];
D.~Bak, Y.~Hyakutake, S.~Kim and N.~Ohta,
arXiv:hep-th/0407253.
















\bibitem{Dab}
A.~Dabholkar,
arXiv:hep-th/0409148.


\bibitem{Sen}
A.~Sen,
arXiv:hep-th/0411255;
V.~Hubeny, A.~Maloney and M.~Rangamani,
arXiv:hep-th/0411272.



\bibitem{Rey}
D.~Bak, S.~Kim and S.~J.~Rey,
arXiv:hep-th/0501014.

\bibitem{Siw}
N.~Ohta, K.~L.~Panigrahi and S.~Siwach,
Nucl.\ Phys.\ B {\bf 674}, 306 (2003)
[arXiv:hep-th/0306186].






\bibitem{Hop}
B.~de Wit, J.~Hoppe and H.~Nicolai,
Nucl.\ Phys.\ B {\bf 305}, 545 (1988).

\bibitem{deW}
B.~de Wit, K.~Peeters and J.~Plefka,
Nucl.\ Phys.\ B {\bf 532}, 99 (1998)
[arXiv:hep-th/9803209].






\bibitem{Tou}
J.~McGreevy, L.~Susskind and N.~Toumbas,
JHEP {\bf 0006}, 008 (2000)
[arXiv:hep-th/0003075].


\bibitem{Ban}
T.~Banks, N.~Seiberg and S.~H.~Shenker,
Nucl.\ Phys.\ B {\bf 490}, 91 (1997)
[arXiv:hep-th/9612157].


\bibitem{HHy}
Y.~Hyakutake and N.~Ohta,
Phys.\ Lett.\ B {\bf 539}, 153 (2002)
[arXiv:hep-th/0204161].


\bibitem{Sad}
D.~Sadri and M.~M.~Sheikh-Jabbari,
Nucl.\ Phys.\ B {\bf 687}, 161 (2004)
[arXiv:hep-th/0312155].








\end{thebibliography}
\end{document}